\documentclass{iopart}
\usepackage{epsfig}

\begin{document}
\title[Quasiparticle entanglement]{Quasiparticle entanglement: redefinition of the vacuum and reduced density matrix approach.} 
\author{P. Samuelsson\dag, E.V. Sukhorukov\ddag, M. B\"uttiker\ddag}
\address{\dag\ Departement of Solid State Theory, Lund University, S\"olvegatan 14 A, S-223 62 Lund, Sweden} 
\address{\ddag\ D\'epartement de Physique Th\'eorique, Universit\'e de Gen\`eve, CH-1211 Gen\`eve 4,
Switzerland} 
\date{\today}

\begin{abstract}
A scattering approach to entanglement in mesoscopic conductors with
independent fermionic quasiparticles is discussed. We focus on
conductors in the tunneling limit, where a redefinition of the
quasiparticle vacuum transforms the wavefunction from a manybody
product state of noninteracting particles to a state describing
entangled two-particle excitations out of the new vacuum. The approach
is illustrated with two examples (i) a normal-superconducting system,
where the transformation is made between Bogoliubov-de Gennes
quasiparticles and Cooper pairs \cite{Sam1}, and (ii) a normal system,
where the transformation is made between electron quasiparticles and
electron-hole pairs \cite{Been1,Sam2}. This is compared to a scheme
where an effective two-particle state is derived from the manybody
scattering state by a reduced density matrix approach.
\end{abstract}
\maketitle 

\section{Introduction.}

Over the last decade, entanglement has come to be viewed as a possible
resource for various quantum information and computation purposes. The
prospect of scalability and integrability of solid state quantum
circuits with conventional electronics has led to great interest in
the investigation of entanglement in solid state systems. A broad
spectrum of proposals for generation, manipulation and detection of
entanglement in solid state systems is given in this volume. Of
particular interest is the entanglement of individual quasiparticles
in mesoscopic conductors. Phase coherence is preserved on long time
scales and over long distances, allowing for coherent manipulation and
transportation of entangled quasiparticles. Moreover, individual
quasiparticles are the elementary entanglable units in solid state
conductors and investigation of quasiparticle entanglement can provide
important insight in fundamental quantum mechanical properties of the
particles and their interactions.

Recently, a number of proposals for creation of entanglement in
mesoscopic systems based on scattering of quasiparticles have been put
forth~\cite{Sam1,Been1,Sam2,Les,Cht,Faoro,Been2,Sam3,Been3,Leb1,Leb2,Naz}. Our
main interest here is to discuss a central aspect of such systems
operating in the tunneling regime, namely the role of the redefinition
of the vacuum in creating an entangled two-particle state. To this aim
we consider the two original proposals, Refs. \cite{Sam1} and
\cite{Been1,Sam2}, were the ground state reformulation was
discussed. The emphasis in these works was on entanglement of the
orbital degrees of freedom~\cite{Sam1}, the discussion however applies
equally well to spin entanglement. In the first proposal~\cite{Sam1},
we investigated a normal mesosocopic conductor contacted to a
superconductor. The superconductor was treated in the standard
mean-field description, giving rise to a Bogoliubov-de Gennes
scattering picture with independent electron and hole
quasiparticles. It was shown that Andreev reflection at the
normal-superconducting interface together with the redefinition of the
vacuum can gives rise to an entangled two-electron state emitted from
the superconductor into the normal conductor. Second, Beenakker {\it
et al}~\cite{Been1} and later the authors~\cite{Sam2} investigated a
normal conductor in the quantum Hall regime. It was shown that the
scattering of individual electron quasiparticles together with the
redefinition of the vacuum can give rise to emission of entangled
electron-hole pairs from the scattering region.

In the present paper, first a general framework for entanglement of
independent fermionic quasiparticles in mesoscopic conductors is
presented. The role of the system geometry and the accessible
measurements in dividing the conductor into subsystems as well as
defining the physically relevant entanglement is emphasized. We then
consider a simple, concrete example with a multiterminal beamsplitter
geometry and derive the emitted manybody scattering state. Two
different approaches to the experimentally accessible two-particle
entanglement are discussed. First, a general scheme for arbitrary
scattering amplitudes based on a reduced density matrix approach is
outlined and then applied to the state emitted by the beamsplitter
geometry. The orbital entanglement of the reduced state is discussed
in some limiting cases. Second, for the system in the tunneling limit,
we show how an entangled two-particle state is created by a
redefinition of the vacuum. The redefinition leads to a transition
from a single-particle to a two-particle picture, with the
wavefunction being transformed from a many-body state of independent
quasiparticles to a state describing an entangled two-particle state
created out of the redefined vacuum. Based on this discussion, a
detailed investigation of the entanglement in the systems in
Refs. \cite{Sam1,Sam2} is presented.

\section{Entanglement in mesoscopic conductors.}

The concept of entanglement appeared in physics in the mid nineteen
thirties \cite{Schrod} as a curious feature of quantum mechanics giving rise to
strong non-local correlations between spatially separated particles.
The non-local properties of entanglement contradicted a commonly held
"local, realistic" view of nature and led Einstein, Podolsky and Rosen
(EPR) to conclude, in their famous paper \cite{EPR}, that quantum
mechanics was an incomplete theory. With the inequalities of Bell
\cite{Bell}, presented three decades later, it became possible to
experimentally test the predicted non-local properties of entangled
pairs of particles. Since then a large number of experiments have been
carried out, predominantly with pairs of entangled photons
\cite{Gisin,Zeilinger,Aspectnat}, where a clear violation of a Bell
Inequality has been demonstrated, providing convincing evidence
against the local realistic view of nature. To date, however, no
violation of a Bell Inequality with electrons has been demonstrated.

During the last decade, the main interest has turned to entanglement
in the context of quantum information processing \cite{qinfo}. It has
become clear that entanglement can be considered as a resource for
various quantum information tasks, such as quantum cryptography
\cite{quantcryp}, quantum teleportation \cite{quanttelep} and quantum
dense coding \cite{quantcode}. The notion of entanglement as a
resource naturally led to the question of how to quantify the
entanglement of a quantum state. A considerable number of measures of
entanglement have been proposed to date \cite{entrevs1,entrevs2},
ranging from describing abstract mathematical properties of the state
to quantifying how useful the state is for a given quantum information
task.
\begin{figure}[h]   
\centerline{\psfig{figure=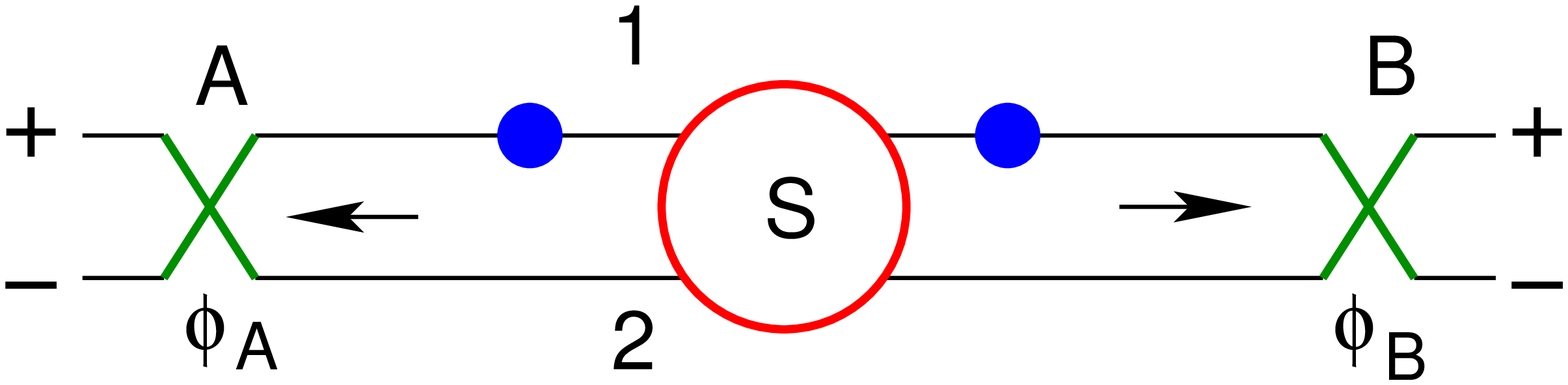,width=10.0cm}}
\centerline{\psfig{figure=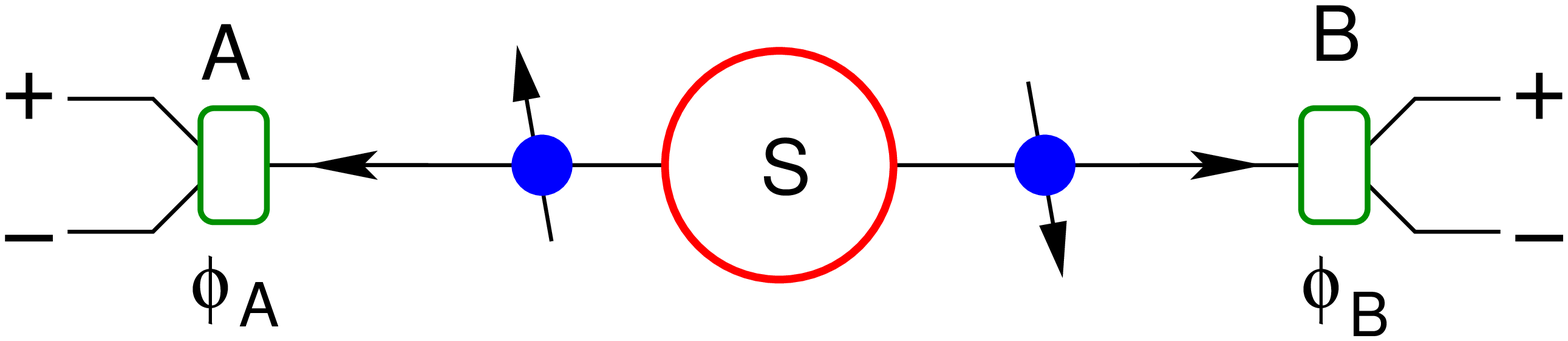,width=10.0cm}}
\caption{Picture of orbital (upper) and spin (lower) entanglement
scheme. From the source S, an orbitally entangled, e.g.
$|\tilde \Psi_O\rangle=1/\sqrt{2}(|1\rangle_A|1\rangle_B+|2\rangle_A|2\rangle_B)$,
or spin entangled, e.g.
$|\tilde\Psi_S\rangle=1/\sqrt{2}(|\uparrow\rangle_A|\downarrow\rangle_B-|\downarrow\rangle_A|\uparrow\rangle_B)$
state is emitted, with one particle propagating towards $A$ and one
towards $B$. At $A$ and $B$, the particles are modulated, e.g. via single qubit rotations parameterized by angles $\phi_A$ and $\phi_B$. The particles are then detected in electronic reservoirs $+$ and $-$.}
\label{figents}
\end{figure}
It is mainly the prospect of quantum information processing in solid
state conductors that has motivated the recent interest in
entanglement in mesoscopic conductors. In this work we do however not
try to answer the ambitious question what quantum information tasks
can be performed in mesoscopic systems. We do also not consider any
particular measure of entanglement, instead we focus the discussion on
the entangled quantum state. Given the state, the entanglement,
quantified by ones measure of liking, can in principle be
calculated. Therefore we are interested in the basic first step in a
quantum information processing scheme, namely the creation,
manipulation and detection of quasiparticle entanglement. A large
number of implementations of this first step have been proposed,
typically a certain mechanism is suggested that leads to emission of
entangled particles from a "source" S. The particles propagate out to
spatially separated regions A and B, where they are manipulated and
detected. Two schematics of generic systems for orbital and spin
entanglement are shown in Fig. \ref{figents}.

Very recently, in a number of works
\cite{Sam1,Been1,Sam2,Les,Cht,Faoro,Been2,Sam3,Been3,Leb1,Leb2}
entanglement in systems of quasiparticles has been investigated within
the framework of scattering theory. These proposals are of particular
interest because working with independent particles allows for a
complete characterization of the emitted many-body state for arbitrary
scattering amplitudes. Moreover, the notion of entanglement without
direct interaction between the quasiparticles, a known concept in the
theory of entanglement \cite{Vedral}, has appeared puzzling to members
of the mesoscopic community. The perception in the mesoscopic physics
community has started to change only with the apperance of
Ref. \cite{Been1}.  This makes it important to thoroughly analyze the
origin of the entanglement. Here we contribute to such an analysis,
focusing on the creation of entangled two-particle states due to
redefinition of the vacuum. For comparison, an approach based on the
reduced two-particle density matrix is discussed as well. Although not
investigated here, it is probable that the two types of states turn
out to be of different ``usefulness'' for quantum information
processing, making a detailed comparision of interest. We start by
stating some important general properties of entanglement and comment
on their application to entanglement in mesoscopic conductors.

i) The entanglement of a state in a given system depends on how the
system is formally parted in to subsystems~\cite{Peres}. If the system
is considered as one entity, i.e. with all quasiparticles living in
the same Hilbert space, one has to consider the question of
entanglement of indistinguishable
particles~\cite{Schliemann1,Schliemann2}. For a mesoscopic system of
independent fermionic quasiparticles, the ground state is given by a
product state in occupation number formalism, i.e. the wavefunction is
a single Slater determinant. Apart from the correlations due to
fermionic statistics the particles show no correlations and the state
is not entangled. However, if the system is considered as consisting of
several spatially separated subsystems, one can pose questions about
the entanglement between spatially separated, distinguishable
quasiparticles living in the different subsystems. The latter is
typically the case in the proposed mesoscopic systems (see
Fig. \ref{figents}), which are naturally parted into the source S and
the two regions A and B. The entanglement between two particles, one
in A and one in B, is in many situations nonzero. Quite generally, the
physically relevant partitions into subsystems are defined by the
system geometry and the possible measurements one can perform on the
system~\cite{Zanardi,Zanardi2}.

ii) The entanglement that can be investigated and quantified is
determined by the possible measurements one can perform on the
system. Although a wide variety of measurements in mesoscopic
conductors in principle can be imagined, the most commonly measured
quantities are currents and current correlators, i.e. current noise
\cite{Butt2,Buttrev}. In particular, in several recent works it was
proposed to detect entanglement via measurements of current
correlators
\cite{Sam1,Been1,Sam2,Cht,Faoro,Been2,Sam3,Been3,Leb1,Leb2,Naz,Sukh,Oliver}. This
leads us to focus the discussion on entanglement detectable with
current correlation measurements. All current cross correlation
measurements, needed to investigate spatially separated particles, are
to date measurements of second order correlators. Importantly, the
second order current correlators are two-particle observables and can
thus only provide direct information about two-particle
entanglement. We thus limit our investigations here to the
two-particle properties of the state \cite{Carlocomm}.

Considering the typical mesoscopic entanglement setup in
Fig. \ref{figents}, the points i) and ii) lead us to discuss the
entanglement between two spatially separated particles, one in A and
one in B, detectable via correlations of currents flowing out into the
reservoirs. Such a bipartite system is also the most commonly
considered one in investigations of entanglement. Importantly, the
entanglement can be both in the orbital \cite{Sam1} as well as in the
spin degrees of freedom, the relevant entanglement depends on the
proposed system geometry which can be designed to investigate orbital
\cite{Sam1,Been1,Sam2,Been2} or spin
\cite{Les,Cht,Faoro,Leb1,Leb2,Naz,Sukh,Sam4} entanglement. We also
note that our measurement based scheme excludes all types of
occupation number or Fock-space entanglement, e.g. the linear
superposition of two particles at $A$ and two at $B$ [which in an
occupation number notation can be written as the Fock-space entangled
state $|0\rangle_A|2\rangle_B+|2\rangle_A|0\rangle_A$]. There is thus
no need to enforce additional constraints~\cite{Wiseman1,Wiseman2} on
the state to exclude the physically irrelevant Fock-space
entanglement.

\section{System and scattering state.}

To clearly illustrate the basic principles, we present here the
formalism for independent quasiparticles in the most elementary
mesoscopic system possible. We focus the discussion on the orbital
entanglement which has the advantage that it can be manipulated and detected \cite{Sam1,Sam2} with
existing experimental techniques. However, for completeness, spin information is
retained throughout the discussion.
\begin{figure}[h]   
\centerline{\psfig{figure=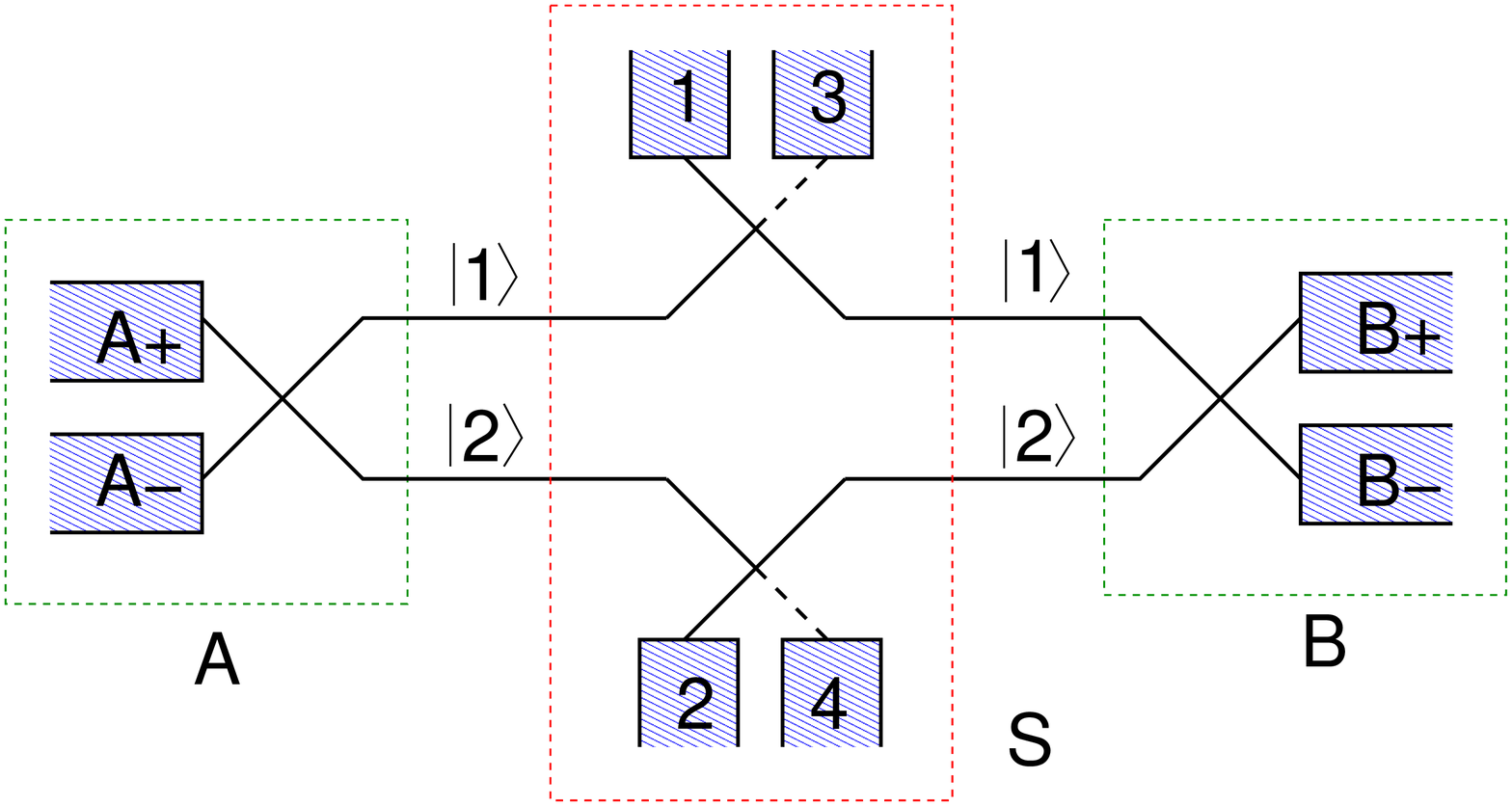,width=12.0cm}}
\caption{Schematic of the entanglement setup. The source region $S$
(dashed red box) consists of two beamsplitters connected to four
reservoirs $1,2,3$ and $4$. The regions $A$ and $B$ (dashed blue boxes)
consist of one beamsplitter and two reservoirs each. The orbital
modes $|1\rangle$ and $|2\rangle$ out of the source region are displayed. Reservoirs
$1$ and $2$ are kept at a potential $V$, the remaining reservoirs are
grounded.}
\label{figcond}
\end{figure}

The system is shown in Fig. \ref{figcond}. In the source region, two
single mode reflectionless beamsplitters are connected to four
electronic reservoirs $1,2,3$ and $4$. A voltage bias $V$ is applied
to reservoirs $1$ and $2$, while reservoirs $3$ and $4$ are kept at
zero bias. Quasiparticles injected from reservoirs $1$ and $2$ scatter
at the beamsplitters and propagate out towards regions $A$ and $B$.
In $A$ and $B$, the quasiparticles are scattered at another pair of
beamsplitters (i.e. a local single qubit rotation) and then detected
in four reservoirs $A+,A-,B+$ and $B-$, all kept at zero potential. We
emphasize that all components of this system can be realized in a
conductor in the quantum Hall regime \cite{Sam2}.

In this work we are interested in the entanglement of the state
emerging from the source region, {\it before} the quasiparticles reach
the beamsplitters in regions $A$ and $B$. The manipulation and
detection processes taking place in $A$ and $B$ are thus not
investigated. The two orbital modes $|1\rangle$ and $|2\rangle$, with
labels $1$ and $2$ denoting from which reservoir the particles are
emerging (see Fig. \ref{figcond}), constitute the orbital two-level,
or pseudo-spin, system. Working within the quasiparticle scattering
approach \cite{Buttrev}, we introduce operators
$a_{m\sigma}^{\dagger}(E)$ creating electron quasiparticles at energy
$E$ with spin $\sigma$, incident from reservoirs $m=1,2,3$ and
$4$. The energy is counted from the Fermi level of the unbiased
reservoirs. The transport state of the system at zero temperature is
given by
\begin{equation}
|\Psi\rangle=\prod_{0<E<eV}a_{1\uparrow}^{\dagger}(E)a_{1\downarrow}^{\dagger}(E)a_{2\uparrow}^{\dagger}(E)a_{2\downarrow}^{\dagger}(E)|0\rangle
\label{gs}
\end{equation}
where $|0\rangle$ is the quasiparticle vacuum, a filled Fermi sea at
energies $E<0$. This describes filled, noiseless streams of electrons
emitted from reservoirs $1$ and $2$, propagating towards the first
pair of beamsplitters. To describe the state after scattering at the
beamsplitters, we introduce operators $b_{Am'\sigma'}(E)$ and
$b_{Bm'\sigma'}(E)$ for particles originating from reservoir $m'$,
propagating from the beamsplitters towards regions $A$ and $B$
respectively. The operators $b_{Am'\sigma'}(E)$ and
$b_{Bm'\sigma'}(E)$ are related to $a_{m\sigma}(E)$ via the scattering
matrices of the beamsplitters. For simplicity we consider identical
beamsplitters, taken independent on $E$ and $\sigma$, giving
(suppressing spin and energy index)
\begin{equation}
\left( \begin{array}{c} b_{A1} \\ b_{B1}\end{array}\right)=\left( \begin{array}{cc} r & t' \\ t & r' \end{array}\right)\left( \begin{array}{c} a_{1} \\ a_{3}\end{array}\right), \hspace{0.1cm} \left( \begin{array}{c} b_{A2} \\ b_{B2}\end{array}\right)=\left(\begin{array}{cc} r & t' \\ t & r' \end{array}\right)\left( \begin{array}{c} a_{2} \\ a_{4}\end{array}\right)
\label{scatmat}
\end{equation}
We note that only static scatterers are considered here, the
discussion could however straightforwardly be extended to time
dependent scatterers, recently investigated in the context of
entanglement in Refs. \cite{Sam6,Been4}. 

To simplify the notation, a collective quantum number
$n=\{1\uparrow,1\downarrow,2\uparrow,2\downarrow\}$ denoting both
orbital mode and spin is introduced. The state describing particles
propagating out towards $A$ and $B$ can then be written
\begin{eqnarray}
|\Psi\rangle&=&\prod_{0<E<eV,n}\left[rb_{An}^{\dagger}(E)+tb_{Bn}^{\dagger}(E)\right]|0\rangle
\label{psiout}
\end{eqnarray}
Due to the scattering, the outflowing streams are noisy and the
properties of the particles flowing towards $A$ and $B$ can be
investigated via noise measurements. Here we are interested in the
entanglement between two particles in the outflowing streams, one
towards $A$ and one towards $B$. Importantly, the state $|\Psi\rangle$
in Eq. (\ref{psiout}) is a many-body state, it describes a linear
superposition of different number of particles at $A$ and $B$, ranging
from zero to in principle infinity. Since only two-particle
entanglement is considered, one thus needs to deduce the two-particle
properties of the state $|\Psi\rangle$. Below we present two different
approaches to do this, giving rise to two different quantum states
with in general different entanglement.

\section{Reduced two-particle density matrix approach.}

We first discuss a general approach, applicable for arbitrary
scattering amplitudes. Starting with the properties of the correlators
of outflowing currents towards A and B, we note that the cross
correlators (in the most general situation) are determined by averages
of the type \cite{Buttrev}
\begin{eqnarray}
&&\langle b_{An}^{\dagger}(E)b_{Am}(E')b_{Bk}^{\dagger}(E'')b_{Bl}(E''')\rangle \nonumber \\
&=&\langle b_{An}^{\dagger}(E)b_{Bk}^{\dagger}(E'')b_{Bl}(E''')b_{Am}(E')\rangle\propto\rho_{nm}^{kl}(E,E',E'',E''')
\label{twopart}
\end{eqnarray}
using in the second step the anticommutation relations for fermionic
operators. The term $\rho_{nmkl}(E,E',E'',E''')$ is by definition an
element of the reduced two-particle density matrix (in the energy
basis). The current correlators are thus fully characterized by the
reduced two-particle density matrix
\begin{eqnarray}
\rho&=&\sum_{nmkl}\int dEdE'dE''dE''' \rho_{nm}^{kl}(E,E',E'',E''') \nonumber \\
&\times& b^{\dagger}_{An}(E)b_{Bk}^{\dagger}(E'')|0\rangle\langle0|b_{Bl}(E''')b_{Am}(E').
\label{densmat}
\end{eqnarray}
Accordingly, the entanglement potentially detectable via noise
measurements is the entanglement of the reduced two-particle density
matrix. Clearly, this is the situation for any two-particle
observable. It should be emphasized that the two-particle density
matrix physically describes the correlations of two particles out of
the streams flowing towards $A$ and $B$, leaving all other particles
unobserved. This is qualitatively different from a projection of a
two-particle state out of the full many-body state, where only the
components of the state containing exactly two particles are selected.
It is shown below that these two different ways of extracting a
two-particle state out of a many particle state can give rise to
different states, and consequently to different entanglement.

Even with the object of interest confined to the reduced two-particle
density matrix, a full characterization of the entanglement is
cumbersome since $\rho$ is in general a mixed state and the two
particles live in infinite dimensional Hilbert spaces spanned by
$\{E,n\}$. In some simple situations (see e.g. below), the density
matrix in Eq. (\ref{densmat}) can be written as a direct product of
the density matrices for the different degrees of freedom,
e.g. $\rho=\rho_O \otimes \rho_S \otimes \rho_E$, where the subscripts
$O,S$ and $E$ denote orbital, spin and energy respectively. This
allows one to independently characterize the entanglement with respect
to the different degrees of freedom. In particular, the orbital
subspace in Fig. \ref{figcond} and in
e.g. Refs. \cite{Sam1,Been1,Sam2} as well as generically the spin 1/2
space are two-level systems or qubits, giving rise to a system of two
coupled qubits, well studied in the entanglement literature (see
e.g. Ref. \cite{Wooters})

However, in the general case, with arbitrary scattering amplitudes,
the reduced two-particle density matrix can not be written as a direct
product, i.e. $\rho \neq \rho_O \otimes \rho_S \otimes \rho_E$. What
can then be said about the entanglement? Of particular interest is the
entanglement detectable via zero frequency current correlators,
generally the quantity investigated in experiments. The zero frequency
limit effectively projects the operators in Eq. (\ref{twopart}) to the
same energy \cite{Buttrev} . For scattering amplitudes independent on
energy on the scale of $eV$, the energy argument in
Eq. (\ref{densmat}) can be suppressed, giving a reduced density matrix
\begin{eqnarray}
\bar \rho&=&\sum_{nm,kl}\rho_{nm}^{kl}b^{\dagger}_{An}b_{Bk}^{\dagger}|0\rangle\langle0|b_{Bl}b_{Am},\hspace{0.5cm}\rho_{nm}^{kl}\propto\langle b_{An}^{\dagger}b_{Bk}^{\dagger}b_{Bl}b_{Am}\rangle.
\label{densmatOS}
\end{eqnarray}
The density matrix $\bar \rho$ contains information about the orbital
and spin parts of the state only. Interestingly, following the
opposite approach and considering the quasiparticle ($E>0$) correlator
for coincident times, closely related to the electronic counterpart
\cite{Sam2} to the joint detection probability introduced by Glauber
\cite{Glauber} in quantum optics, one finds the same reduced density
matrix $\bar \rho$ as in the zero frequency limit. Similar results for
short time current correlators have been obtained in
Refs. \cite{Cht,Been3,Leb1,Leb2}.

The reduced density matrix $\bar \rho$ for the system under
consideration (see Fig. \ref{figcond}) is evaluated from
Eqs. (\ref{gs}), (\ref{scatmat}) and (\ref{densmatOS}) to be
\begin{eqnarray}
\bar \rho=\frac{1}{8}\left({\bf 1}_O\otimes {\bf 1}_S-F_O\otimes F_S\right),\hspace{0.5cm}F=\left(\begin{array}{cccc} 1&0&0&0 \\ 0&0&1&0 \\ 0&1&0&0 \\ 0&0&0&1\end{array}\right)
\label{sodens}
\end{eqnarray} 
expressed in the orbital basis $\{|1\rangle_A|1\rangle_B,
|1\rangle_A|2\rangle_B,|2\rangle_A|1\rangle_B,|2\rangle_A|2\rangle_B\}$
and the spin basis $\{|\uparrow\rangle_A|\uparrow\rangle_B,
|\uparrow\rangle_A|\downarrow\rangle_B,|\downarrow\rangle_A|\uparrow\rangle_B,|\downarrow\rangle_A|\downarrow\rangle_B\}$. Here
${\bf 1}$ is the $4\times 4$ unit matrix. Note that since the
particles in $A$ and $B$ are distinguishable there is no need for
anti-symmetrization and we can use the notation e.g.
$a_{A1\uparrow}^{\dagger}a_{B2\downarrow}^{\dagger}|0\rangle=|1\rangle_A|2\rangle_B\otimes|\uparrow\rangle_A|\downarrow\rangle_B$. Formally,
the first term in Eq. (\ref{sodens}) results from the direct pairing
of the operators in the bracket in Eq. (\ref{densmat}) and is given by
the direct product of the single particle density matrices at $A$ and
$B$. The second term results from the exchange pairing. Since the
direct term by definition does not describe any nonlocal correlations
(it is a separable state with respect to $A$ and $B$), it is the
exchange correlations which are responsible for the entanglement. For
the current correlators, it should be noted that the first term
determines the product of the averaged currents, while the second term
determines the irreducible current correlators, i.e. the correlators
of the current fluctuations, the noise. We also note that $\bar \rho$
is independent on scattering amplitudes, a consequence of the
assumption of identical, spin independent beamsplitters.

As is clear from Eq. (\ref{sodens}), even though we consider a simple
geometry with spin-independent scattering, the reduced density matrix
$\bar \rho$ is not a direct product between orbital and spin part,
i.e. $\bar\rho \neq \rho_O \otimes \rho_S$. Many observables are
however not sensitive to the spin degree of freedom, as is the case
for cross correlators between total currents
$I=I_{\uparrow}+I_{\downarrow}$, typically the cross correlators
investigated in mesoscopic conductors. The effective orbital density
matrix accessible via current correlators is then obtained by tracing
$\bar \rho$ over the spin degree of freedom, giving
\begin{eqnarray}
\rho_O\equiv\mbox{tr}_S\left[\bar \rho\right]=\frac{1}{6}\left(2{\bf
1}-F\right)=\frac{1}{6}\left({\bf 1}+2|\Psi_O\rangle
\langle\Psi_O|\right)
\label{odens}
\end{eqnarray} 
with
$|\Psi_O\rangle=1/\sqrt{2}\left[|1\rangle_A|2\rangle_B-|2\rangle_A|1\rangle_B\right]$. This
state, an example of a Werner state \cite{Werner}, can via suitable
local transformations be written on a separable form
\cite{Entcrit1,Entcrit2} with respect to $A$ and $B$ and is
consequently not entangled. We note that the same holds for the
reduced spin density matrix $\rho_S$, obtained by tracing over orbital
degrees of freedom.

A different situation occurs if one considers a spin-polarized system,
as was done e.g. in \cite{Been1,Sam2}. In this case the density matrix
is purely orbital, obtained from Eq. (\ref{densmatOS}) by suppressing
the spin notation. For the system in Fig. \ref{figcond} we obtain
\begin{eqnarray}
\rho_O=\frac{1}{2}\left({\bf
1}-F\right)=|\Psi_O\rangle \langle\Psi_O|,\hspace{0.5cm}
\label{eeent}
\end{eqnarray}
an orbital singlet, i.e. a maximally entangled state, again
independent on scattering amplitudes. This result can be understood by
considering the energy and spin independent incoming two-particle
state $|\Psi\rangle=a_{1}^{\dagger}a_{2}^{\dagger}|0\rangle$, the
version of the state in Eq. (\ref{gs}) appropriate under the stated
assumptions. The corresponding outgoing state is
\begin{eqnarray}
|\Psi\rangle=\left[r^2b_{A1}^{\dagger}b_{A2}^{\dagger}+t^2b_{B1}^{\dagger}b_{B2}^{\dagger}+rt\left(b_{A1}^{\dagger}b_{B2}^{\dagger}-b_{A2}^{\dagger}b_{B1}^{\dagger}\right)\right]|0\rangle
\label{twoelstate}
\end{eqnarray}
The first two terms describe two particles at A or two at B, while the
last term describes one particles at A and one at B. As is clear from
Eq. (\ref{densmatOS}), only the last term, which is just
$|\Psi_O\rangle$, contributes to $\rho_O$. Importantly, for the
two-particle state in Eq. (\ref{twoelstate}), the reduced density
matrix approach gives the same result as projecting out the part of
the state which contains one particle at $A$ and one at $B$. Both
procedures are thus equivalent to a post-selection of entanglement, as
originally discussed in quantum optics \cite{ShiAlley} (see
e.g. Ref. \cite{Bose} for a discussion for fermions). Various issues
of projection and post-selection were recently discussed in a number of
works on entanglement in mesoscopic conductors with arbitrary
scattering amplitudes \cite{Sam2,Been2,Leb1,Leb2} (for a related
discussion, see also \cite{Faoro}).

It is interesting to note the clear difference between an orbital
state obtained by tracing $\bar \rho$ over the spin degrees of freedom
and the orbital state in a spin polarized system. The difference can
be attributed to the fact that only spins of the same spieces are
nonlocally correlated, i.e. contribute to the entanglement of the
state. Detecting (without spin resolution) two particles, one at $A$
and one at $B$, the probability of obtaining two identical spins is
only one half, reducing the entanglement of the state to zero.

\section{Tunneling limit, vacuum redefinition.}

A qualitatively different approach to the characterization of the emitted state can be
taken in the limiting case of a tunneling system, as was done in
Refs. \cite{Sam1,Been1,Sam2}.
\begin{figure}[h]   
\centerline{\psfig{figure=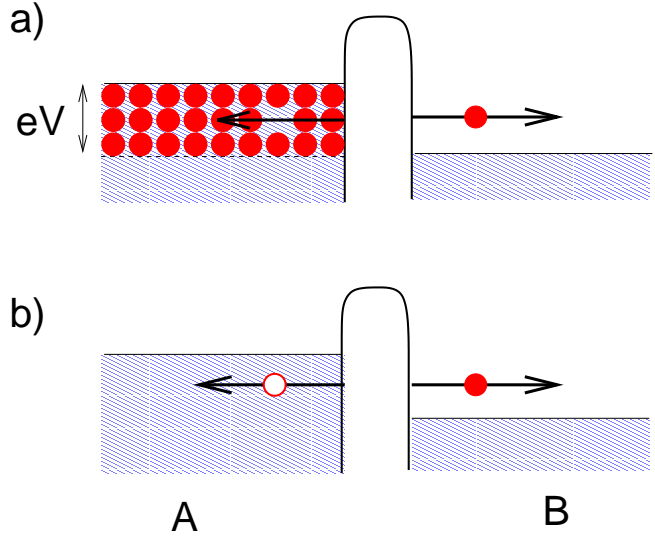,width=9.0cm}}
  \caption{Pictures for particles flowing out from a beamsplitter in
the source region. a) Single particle scattering picture. A
transmission of a particle in the filled stream incident on the
beamsplitter leads to creation of an electron flowing out towards $B$
and a missing electron, i.e. a hole, in the streams flowing out towards
$A$. b) Electron-hole pair creation picture, the single particle
scattering event creates an electron-hole pair on top of the redefined
vacuum, with the electron moving out towards $B$ and the hole towards
$A$.}
\label{fig3}
\end{figure}
Consider the state in Eq. (\ref{psiout}). In the tunneling limit, $t \ll 1$,
we can expand the product to first order in $t$ as

\begin{eqnarray}
\prod_{0<E<eV,n}\left[rb_{An}^{\dagger}(E)+tb_{Bn}^{\dagger}(E)\right]
\nonumber \\
\lo =...\left[rb_{An}^{\dagger}(E)+tb_{Bn}^{\dagger}(E)\right]
\left[rb_{An'}^{\dagger}(E)+tb_{Bn'}^{\dagger}(E)\right]...
\nonumber \\
\lo \times\left[rb_{An}^{\dagger}(E')+tb_{Bn}^{\dagger}(E')\right]...
\left[rb_{An}^{\dagger}(E'')+tb_{Bn}^{\dagger}(E'')\right]...\nonumber\\
\lo =\prod_{0<E<eV,n}b_{An}^{\dagger}(E)+t\int_0^{eV} dE'\sum
_{n'}
..b_{An'}^{\dagger}(E)..b_{Bn'}^{\dagger}(E')..b_{An}^{\dagger}(E'')..
\label{expand}
\end{eqnarray}
where the last term contains $b_{An}$ operators at all energies and
$n$'s except $\{E',n'\}$, in the same order as in the line above (we
consider a continuous spectrum). Using the property that
$b_{An'}(E')b_{An'}^{\dagger}(E')|0\rangle=|0\rangle$ and that the
operator products $b_{An'}(E')b_{An'}^{\dagger}(E')$ and
$b_{Bn'}(E')^{\dagger}b_{An'}(E')$ commutes with all $b$-operators at
different energies $E\neq E'$ or $n\neq n'$, we can write the state in
Eq. (\ref{psiout}) in a suggestive way by inserting
$b_{An'}(E')b_{An'}^{\dagger}(E')$ in front of $|0\rangle$ and
reordering the operators, giving
\begin{eqnarray}
|\Psi\rangle&=&\prod_{0<E<eV,n}b_{An}^{\dagger}(E)|0\rangle \nonumber \\
&+& t\int_0^{eV} dE'\sum_{n'} b_{Bn'}^{\dagger}(E')b_{An'}(E') \prod_{0<E<eV}b_{An}^{\dagger}(E)|0\rangle \nonumber \\
&=&\left[1+t\int_0^{eV} dE' \sum_{n'} b_{Bn'}^{\dagger}(E')b_{An'}(E')\right]\prod_{0<E<eV,n}b_{An}^{\dagger}(E)|0\rangle
\label{wavefcn}
\end{eqnarray}
The first term describes filled streams of quasiparticles flowing out
towards $A$. The second term, with a small amplitude $t\ll 1$,
describes the same filled streams with one electron missing and in
addition a second electron propagating out towards $B$ (in mode $1$ or
$2$). It is thus very natural to incorporate the filled, noiseless
stream of quasiparticles flowing out towards $A$ into a new vacuum
$|\bar 0\rangle$, the second term then describes an electron-hole
excitation out of the redefined vacuum. Formally, we can write the new
vacuum in terms of the old vacuum $|0\rangle$ as
\begin{eqnarray}
|\bar 0 \rangle=\prod_{0<E<eV,n}b_{An}^{\dagger}(E)|0\rangle
\end{eqnarray}
The new ground state is thus a filled Fermi sea at energies $E<0$ in
$B$ and at $E<eV$ in $A$. The fermionic operators describing
excitations out of the new vacuum are defined as
\begin{eqnarray}
c^{\dagger}_{A\bar n}(E')=\pm b_{An}(E), \hspace{0.5cm}
c^{\dagger}_{Bn}(E)=b_{Bn}^{\dagger}(E)
\label{cop}
\end{eqnarray}
where $\bar n$ denotes a spin opposite to $n$, $+(-)$ is for spin up
(down) in $\bar n$ and $E'=eV-E$. The transformation in
Eq. (\ref{cop}) is thus equivalent to an inverse Bogoliubov
transformation. We can then write the wavefunction in
Eq. (\ref{wavefcn}) as (reintroducing spin and orbital notation) 
\begin{eqnarray}
|\Psi\rangle=|\bar 0\rangle+|\bar \Psi \rangle
\end{eqnarray}
with
\begin{eqnarray} 
|\bar \Psi \rangle=t\int_0^{eV} dE\sum_{j=1,2} \left[c_{B\downarrow j}^{\dagger}(E)c_{A\uparrow j}^{\dagger}(E')-c_{B\uparrow j}^{\dagger}(E)c_{A\downarrow j}^{\dagger}(E')\right]|\bar 0\rangle 
\label{wavefcn2}
\end{eqnarray}
The wavefunction $|\bar \Psi \rangle$ describes a two-particle
excitation, an electron-hole pair, out of the redefined vacuum $|\bar
0 \rangle$. The redefinition of the vacuum thus gives rise to a
transformation from a {\it picture with a many-body state of independent
particles to a picture with two-particle excitations out of a ground
state}.

There are a number of important conclusions to be drawn from the
result above, and to be put in relation to the result for the reduced density
matrix approach:

(i) The state is wavepacket-like, i.e. it consists of a sum of
electron-hole pairs at different energies, a detailed characterization
of a similar wavepacket state emitted in a normal-superconductor
system is given by us in Ref. \cite{Sam4}. As pointed out in
Ref. \cite{Sam1}, the average time between two subsequent wavepackets
is much longer than the ``width'' of each wavepacket, i.e. subsequent
entangled pairs are well separated in time, in contrast to the reduced
two-particle state in the large transparency limit. This temporal
separation probably makes the entangled state created by a
redefinition of the vacuum more useful for quantum information
processing, due to the possibility of addressing the individual
entangled pairs. We remark that the formal procedure used above to
calculate the state in Eq. (\ref{wavefcn2}) is essentially identical
to the one for calculating the wavefunction of the two-photon state
emitted in a parametric down conversion process in optics \cite{Ou}.

(ii) In a first quantization notation, we can write the wavefunction
of the excitation 
\begin{eqnarray}
|\bar \Psi \rangle&=&t\int_0^{eV}
dE\left[|1\rangle_A|1\rangle_B+|2\rangle_A|2\rangle_B\right]\nonumber
\\
&\otimes&\left[|\downarrow\rangle_A|\uparrow\rangle_B-|\uparrow\rangle_A|\downarrow\rangle_B\right]\otimes|E'\rangle_A|E\rangle_B
\label{ehstate}
\end{eqnarray}
This state is a direct product of the orbital, spin and energy parts
of the state, in contrast to the general situation for the reduced
two-particle state. Both the orbital and the spin state are maximally
entangled Bell states. The spin state is a singlet as one would expect
of an excitation out of a spinless groundstate, created by
spin-independent scattering.

(iii) As was emphasized in Ref. \cite{Been1}, the redefinition of the
vacuum is possible only in fermionic systems, i.e. it relies on the
existence of a filled Fermi sea such that a removal of an electron
below the Fermi energy creates a hole quasiparticle. This is further
emphasized by the fact that the new groundstate, just as the initial
one is noiseless.

(iv) The correlators between electron currents are simply related to
the correlators between electron and hole currents as $\langle \Delta
I^e(t)\Delta I^h(t')\rangle=-\langle \Delta I^e(t')\Delta
I^e(t)\rangle$. Consequently, the electron-hole correlators are
experimentally accessible and the electron-hole entanglement is, in
line with the discussion above, a physically relevant object to
study.

Importantly, the redefinition of the vacuum and the transformation to
an electron-hole picture can be performed for an arbitrary
transmission. However, in this case the resulting state describes a
superposition of different numbers of electron-hole pairs.  To obtain
a two-particle state, one has to calculate a reduced electron-hole
density matrix along the same line as in Eq. (\ref{densmat}),
i.e. replacing the electron operator $a_{An}^{\dagger}$ with the
quasiparticle operator $c_{A\bar n}^{\dagger}$ etc. Performing such a
calculation in the low transparency limit, one obtains as expected
$|\bar \Psi \rangle\langle \bar \Psi|$, with $|\bar \Psi \rangle$
given by Eq. (\ref{ehstate}). It is however not possible
from the reduced density matrix approach to conclude whether the
emitted state is a true two-particle state or a reduced two-particle
state.

For arbitrary scattering amplitudes, to quantitatively compare the
reduced density matrix approach for electrons and holes to the results
for electrons discussed above, we consider the simplest situation with
low frequency correlators, i.e. all $c$-operators at equal energy, and
a spin polarized conductor. The reduced orbital density matrix for the
system in Fig. \ref{figcond} is then given by
\begin{eqnarray}
\rho_O^{eh}=\frac{1}{2(1+T)}\left[T{\bf
 1}+2R|\tilde\Psi_O\rangle\langle \tilde \Psi_O|\right], 
\label{ehorbdens} 
\end{eqnarray}
with
$|\tilde\Psi_O\rangle=1/\sqrt{2}\left[|1\rangle_A|1\rangle_B+|2\rangle_A|2\rangle_B\right]$
and the scattering probabilities $R=1-T=|r|^2$. Interestingly, in
contrast to the density matrix in Eq. (\ref{sodens}), the density
matrix $\rho_O^{eh}$ depends on the scattering probabilities. The
state $\rho_O^{eh}$ is a Werner state, entangled for $R>T$. In the
limit $T \ll 1$, one has $\rho_O^{eh}=|\tilde\Psi_O\rangle\langle
\tilde \Psi_O|$, an orbital Bell state, maximally entangled. Moreover,
away from the tunneling limit, the two-particle entanglement in the
electron-hole picture is smaller than the entanglement in the electron
picture [the state in Eq. (\ref{eeent}) is maximally entangled]. This
is in agreement with the findings of Lebedev {\it et al} \cite{Leb1},
who investigated the conditions for a violation of a Bell Inequality
for a scatterer with arbitrary transparency, comparing the electron
and the electron-hole approaches.

The state $\rho_O^{eh}$ can be understood by considering the state in
Eq. (\ref{twoelstate}) after redefining the vacuum and transforming to
an electron-hole picture, giving
\begin{eqnarray}
|\Psi\rangle=\left[r^2+t^2c_{B1}^{\dagger}c_{B2}^{\dagger}c_{A1}^{\dagger}c_{A2}^{\dagger}+rt\left(c_{A1}^{\dagger}c_{B1}^{\dagger}+c_{A2}^{\dagger}c_{B2}^{\dagger}\right)\right]|\bar
 0\rangle
\label{twoehstate}
\end{eqnarray}
The last term is just $|\tilde\Psi_O\rangle$ while the second term,
describing four particles, i.e. two electron-hole pairs, also
contributes to the reduced density matrix in Eq. (\ref{ehorbdens}), it
gives rise to the term $T{\bf 1}$. Interestingly, performing a
projection of $|\Psi\rangle$ onto a state with only two particles, one
in $A$ and one in $B$, one obtains the maximally entangled state
$|\tilde\Psi_O\rangle$, since the terms with two electron-hole pairs
are discarded. The projection approach thus, in the large transparency
limit, overestimates the entanglement detectable via current
correlations.

To illustrate the relevance of the vacuum redefinition, we now discuss
two different systems where this was investigated.

\section{Normal-Superconducting entangler.}

We first consider the case of a normal-superconducting system, to
large extent following Ref. \cite{Sam1}. As in all existing works on
entanglement in normal-superconducting systems
\cite{Les,Cht,Sam4,Recher,Recher2,Bena,Recher3,Bouchiat,Fein}, we
consider the superconductor in the mean-field description. The mean
field Hamiltonian is bilinear in fermionic operators and is
diagonalized by a Bogoliubov transformation, giving rise to a new set
of independent electron and hole like quasiparticles. Importantly,
although the microscopic mechanism for superconductivity is
interaction between electron quasiparticles, in the mean field
description it is again possible to find a picture with noninteracting
quasiparticles. For a noninteracting normal conductor connected to a
superconductor, the whole system can thus be treated within a single
particle scattering approach to the Bogoliubov-de Gennes
equation \cite{Datta}. Consequently, the approach above to the entanglement for
systems of independent quasiparticles can be applied.

\begin{figure}[h]   
\centerline{\psfig{figure=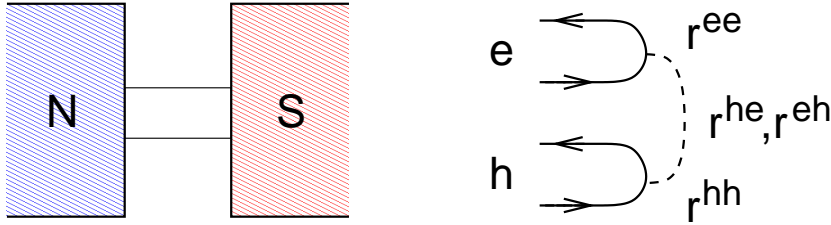,width=11.0cm}}
\caption{Schematic of the conductor. Left: Mesoscopic conductor
contacted to a normal reservoir (blue shaded) and a superconducting
reservoir (red shaded). Right: Paths of scattering particles at
energies well below the superconducting gap. An incoming electron from
the normal reservoir has the amplitude $r^{ee}$ to be reflected as an
electron and $r^{he}$ to be reflected as a hole. For incoming holes,
the corresponding amplitudes are $r^{eh}$ and $r^{hh}$.}
\label{figcondns}
\end{figure}
To illustrate the basic principle of two-particle emission, we first
consider the case of a two-mode (to connect to the orbital discussion
above) normal-superconductor system shown in
Fig. \ref{figcondns}. At the normal-superconductor interface, for
energies well below the superconducting gap, scattering occurs either
as Andreev reflection or as normal reflection. Consider the situation
with a negative bias $-eV$ applied at the normal reservoir while the
superconducting reservoir is kept at zero potential. Counting the
energy from the superconducting chemical potential $\mu_S$, we
consider operators $a_{en}^{\dagger}(E)$ and $a_{hn}^{\dagger}(E)$,
creating electron and hole quasiparticle excitations respectively at
energy $E$ incident from the normal reservoir. The collective quantum
number $n$ denotes as above orbital mode and spin. The state of the
system at zero temperature is given by
\begin{equation}
|\Psi\rangle_{NS}=\prod_{0<E<eV,n}a^{\dagger}_{hn}(E)|0\rangle
\label{gsns}
\end{equation}
describing a filled stream of holes injected from the normal
reservoir, where $|0\rangle$ is the quasiparticle vacuum. The operators
$a_{en}(E)$ and $a_{hn}(E)$ are related to operators $b_{en}(E)$ and
$b_{hn}(E)$ for quasiparticles propagating back from the
normal-superconducting interface towards the normal reservoir via the
scattering matrix (again taken independent on $E$ and $n$) as
\begin{equation}
\left( \begin{array}{c} b_{e} \\ b_{h}\end{array}\right)=\left( \begin{array}{cc} r^{ee} & r^{eh} \\ r^{he} & r^{hh} \end{array}\right)\left( \begin{array}{c} a_{e} \\ a_{h}\end{array}\right)
\end{equation}
The state $|\Psi\rangle_{NS}$ can then be written in terms of the $b$-operators
\begin{eqnarray}
|\Psi\rangle_{NS}&=&\prod_{0<E<eV,n}\left[r^{eh}b_{en}^{\dagger}(E)+r^{hh}b_{hn}^{\dagger}(E)\right]|0\rangle
\label{psioutns}
\end{eqnarray}
Considering the tunneling limit, the Andreev reflection amplitude
$r^{eh}\ll 1$ and one can proceed in exactly the same way as in Eqs. (\ref{expand}) and (\ref{wavefcn}) to arrive at the wavefunction
to leading order in $r^{eh}$ as
\begin{eqnarray}
|\Psi\rangle_{NS}&=&\left[1+r^{eh}\int_0^{eV} dE' \sum_{n'} b_{en'}^{\dagger}(E')b_{hn'}(E')\right]\prod_{0<E<eV,n}b_{hn}^{\dagger}(E)|0\rangle \nonumber \\
\label{wavefcnns}
\end{eqnarray}
The first term describes a filled stream of hole quasiparticles
flowing back towards the normal reservoir. The second term, with a
small amplitude $r^{eh}\ll 1$, describes the same filled stream with
one hole missing and in addition a second electron propagating back.
We can thus proceed as above and incorporate the filled, noiseless
stream of hole quasiparticles flowing out towards the normal conductor
into a new, redefined vacuum $|\bar 0\rangle$, given by
\begin{eqnarray}
|\bar 0 \rangle=\prod_{0<E<eV,n}b_{An}^{\dagger}(E)|0\rangle
\end{eqnarray}
The fermionic operators describing excitations of the new ground state 
are given by the Bogoliubow transformation
\begin{eqnarray}
c^{\dagger}_{n}(E)=b_{en}^{\dagger}(E), \hspace{0.5cm}
c^{\dagger}_{\bar n}(-E)=\pm b_{hn}(E)
\label{copns}
\end{eqnarray}
where $\bar n$ denotes the spin-flip in the electron-hole
transformation with $+(-)$ for spin up (down) in $\bar n$. Note that
the $c$-operators are just standard electron operators. A missing hole
in the filled hole stream is thus just an electron with opposite spin
and energy compared to the superconducting chemical potential $\mu_S$.
\begin{figure}[h]   
\centerline{\psfig{figure=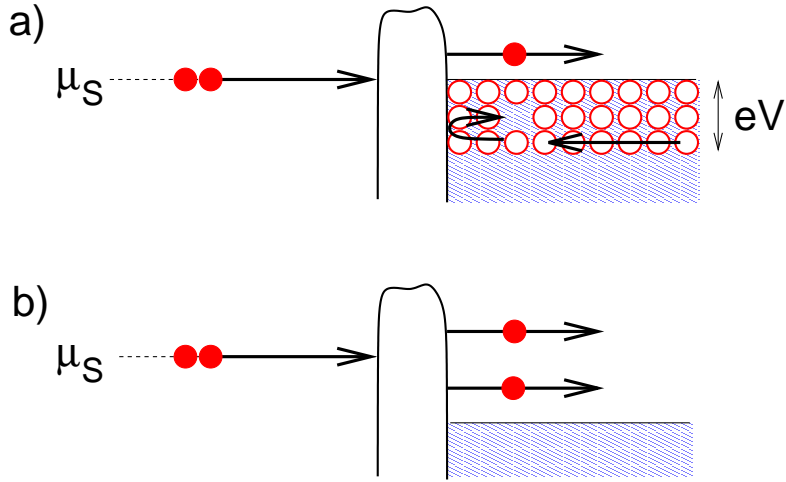,width=11.0cm}}
  \caption{Schematic of a normal-superconducting interface. a)
Bogoliubov-de Gennes picture, a filled stream of holes are incident on
S from N. An Andreev reflection leads to creation of an electron above
the superconducting chemical potential $\mu_S$ and a missing hole in
the back-flowing hole stream. b) Cooper pair tunneling picture, a
Cooper pair tunnels from S to N, leading to a pair of electrons in N,
on top of the ground state.}
\label{fig2}
\end{figure}
We can then write the wavefunction in Eq. (\ref{wavefcn}) as,
reintroducing spin and orbital notation
\begin{eqnarray}
|\Psi\rangle_{NS}=|\bar 0\rangle_{NS}+|\bar \Psi \rangle 
\end{eqnarray}
with 
\begin{eqnarray}
|\bar \Psi\rangle=r^{eh}\int_0^{eV} dE\sum_{j=1,2} \left[c_{j\uparrow}^{\dagger}(E)c_{j\downarrow}^{\dagger}(-E)-c_{j\downarrow}^{\dagger}(E)c_{j\uparrow}^{\dagger}(-E)\right]|\bar 0\rangle
 \label{wavefcn2ns}
\end{eqnarray}
The wavefunction $|\bar \Psi \rangle$ thus describes a two-particle
excitation, an electron pair or Cooper pair, out of the redefined
vacuum $|\bar 0 \rangle$. The redefinition of the vacuum thus gives
rise to a transformation from a picture with a many-body state of
independent particles to a picture with two-electron excitations out
of a ground state. We point out that this approach, shown
schematically in Fig. \ref{fig2}, provides a formal connection between
the scattering and the tunnel Hamiltonian approach (see
e.g. Refs. \cite{Sam4,Recher}).

\begin{figure}[h]
\centerline{\psfig{figure=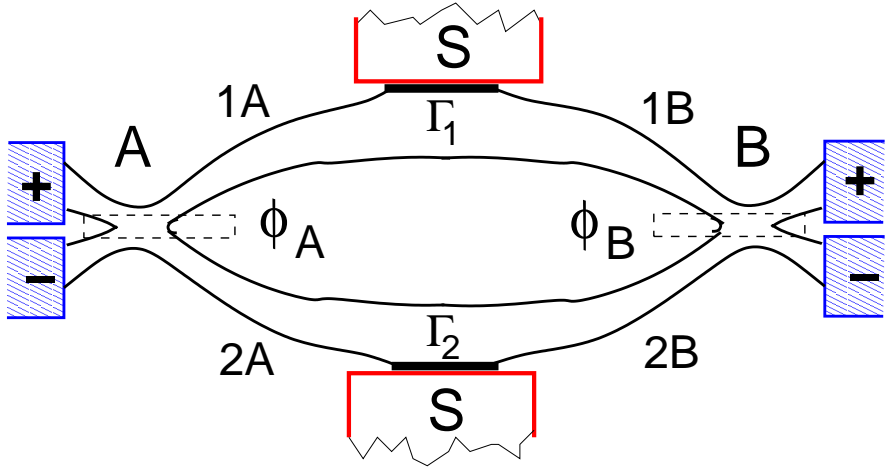,width=0.9\linewidth}}
\caption{Orbital normal-superconductor entangler: A single superconductor (S)
is connected to four normal arms via two tunnel barriers $1$ and $2$
(thick black lines). The arms are joined pairwise in beam splitters A
and B and end in normal reservoirs labeled $+$ and $-$. After Ref. \cite{Sam1}.}
\label{junction}
\end{figure}

To connect this two-particle emission to orbital entanglement, we
consider as a concrete example our proposal in Ref. \cite{Sam1}. The
system geometry is shown in Fig. \ref{junction}, a multiterminal
normal conductor connected via tunnel barriers to a single
superconductor and further to four normal reservoirs. The two regions
at $A$ and $B$, with an electronic beamsplitter and two reservoirs
respectively, constitute the two subsystems. For details of the
proposals we refer the reader to Ref. \cite{Sam1} and
Ref. \cite{Sam5}. To first order in tunnel barrier transparency a pair
of electrons is emitted on top of the new ground state, at
interface $1$ or $2$.
This leads to a state describing a linear superposition of pairs at
$1$ and at $2$. Taking into account that the electrons can be emitted
either towards $A$ or towards $B$, we have the emitted state given by
$|\bar \Psi\rangle_{NS}=|\Psi_1\rangle_{NS}+|\Psi_2\rangle_{NS}$,
where
\begin{eqnarray}
|\tilde \Psi_1\rangle_{NS}&=&r^{eh}\int_{0}^{eV}dE
 \sum_{j=1,2}\sum_{\eta=A,B}
 \left[c_{\eta j \uparrow}^{\dagger}(E)c_{\eta j \downarrow}^{\dagger}(-E)
 \right. \nonumber \\ && \hspace{0.5cm}-
 \left. c_{\eta j\downarrow}^{\dagger}(E)c_{\eta j \uparrow}^{\dagger}(-E)
 \right]|\bar 0\rangle, \nonumber \\
 |\tilde \Psi_2\rangle_{NS}&=&r^{eh}\int_{-eV}^{eV}dE\left[c_{A1\uparrow}^{\dagger}(E)c_{B1\downarrow}^{\dagger}(-E)-c_{A1\downarrow}^{\dagger}(E)c_{B1\uparrow}^{\dagger}(-E)
 \right. \nonumber \\ && +
 \left. c_{A2\uparrow}^{\dagger}(E)c_{B2\downarrow}^{\dagger}(-E)-c_{A2\downarrow}^{\dagger}(E)c_{B2\uparrow}^{\dagger}(-E)\right]|\bar
 0\rangle,
\label{entNSwavefcn12}
\end{eqnarray}
The state $|\tilde \Psi_1\rangle_{NS}$ describes a superposition of
two particles at $A$ and two at $B$. This state does however not
contribute to the noise correlators or, as is clearly the case, to the
two-particle density matrix describing one particle at A and one at
B. The state $|\tilde \Psi_2\rangle_{NS}$ however describes orbitally
(as well as spin-entangled) electron ``wave packet'' pairs, with one
particle at A and one at B. This is the entangled state detected by
the noise.

\section{Normal state entangler.}

We then turn to a normal state entangler working in the Quantum Hall
regime. Such a system was introduced by Beenakker {\it et al} in
Ref. \cite{Been1} and then later considered in \cite{Been2}. In both
Refs. \cite{Been1} and \cite{Been2}, two parallel edgestates connected
via a nonadiabatic, edge channel mixing scattering region was
considered. In Ref. \cite{Sam2} we instead considered a topologically
different Quantum Hall system, a Hanbury Brown Twiss geometry (or
Corbino geometry) with only single edge-states and quantum point
contacts. This highly simplifies the experimental realization of the
proposal. The system is just a proposal for a physical realization of
the system shown in Fig. \ref{figcond}, discussed in detail above, and
we therefore keep the discussion short.

\begin{figure}[t]
\centerline{\psfig{figure=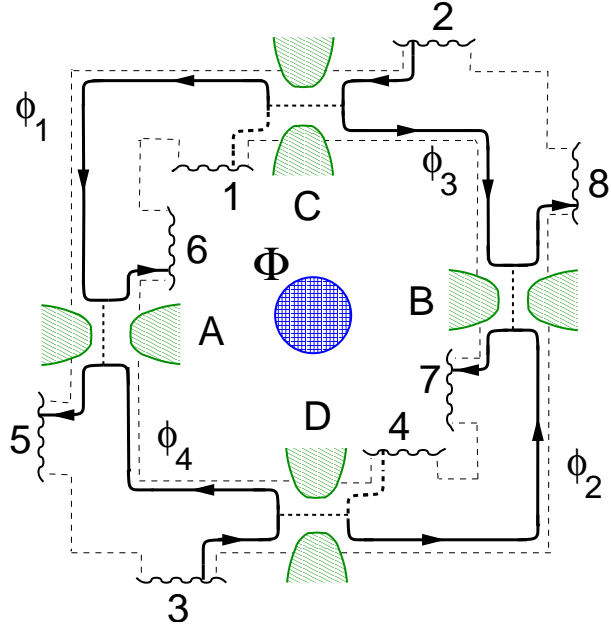,width=8.0cm}}
\caption{Normal-state orbital entangler: a rectangular Hall bar with
inner and outer edges (thin dashed lines) and four quantum point
contacts (grey shaded) with transparencies $T_A,T_B,T_C$ and
$T_D$. Contacts $2$ and $3$ are sources of electrons (a voltage $eV$
is applied against all other contacts which are at ground). Electrons
follow edge states (thick black lines) in the direction indicated by
the arrows. An Aharanov-Bohm flux $\Phi$ penetrates the center of the
sample (shaded). After Ref. \cite{Sam2}.}
\label{figHBT}
\end{figure}

The system is shown in Fig. \ref{figHBT}, for details we refer the
reader to Ref. \cite{Sam2}. A spin-polarized edge state is
considered. We take the transmission and reflection probabilities at
the point contact $C$ to be $T_C=1-R_C=T$ and at $D$ to be
$T_D=1-R_D=R$. After scattering at $C$ and $D$, the state $|\Psi
\rangle$ consists of two contributions in which the two particles fly
off one to $A$ and one to $B$, and of two contributions in which the
two particles fly both off towards the same detector QPC.
Consider now the case of strong asymmetry $R \ll 1$, where almost no
electrons are passing through the source QPC's towards $B$. Performing
the reformulation of the ground state and the transformation ta an
electron-hole picture, we can directly write the full state $|\Psi
\rangle$ to leading order in $\sqrt{R}$ as $|\Psi\rangle=|\bar
0\rangle+\sqrt{R}|\tilde \Psi\rangle$, with
\begin{equation}
|\tilde \Psi \rangle=\int_0^{eV}\hspace{-0.3cm}dE
\left[c_{3B}^{\dagger}(E)c_{3A}^{\dagger}(E)-c_{2B}^{\dagger}(E)c_{2A}^{\dagger}(E)\right]|\bar
0\rangle
\label{newgs}
\end{equation}
Due to the redefinition of the vacuum, we can interpret the resulting
state $|\tilde \Psi \rangle$ as describing a superposition of
"wavepacket"-like electron-hole pair excitations out of the new
vacuum, i.e. an orbitally entangled pair of electron-hole
excitations. This is just equivalent to the result in
Eq. (\ref{ehstate}).

\section{Conclusions}
We have presented a scattering approach to entanglement in mesoscopic
conductors with independent fermionic quasiparticles. The role of the
system geometry and accessible measurements in defining the relevant
entanglement was discussed. As a simple example, a multiterminal
mesoscopic conductor with spin-independent scattering was investigated
in detail, deriving an expression for the manybody scattering state
emitted by the conductor. The focus was on orbital entanglement,
accessible with present day experimental technics.

Two different approaches to a two-particle state were considered. The
main focus was on conductors in the tunneling limit, where entangled
two-particle states arise from a redefinition of the quasiparticle
vacuum. The redefinition of the vacuum leads to a transition from a
picture with a many-body state of independent particles to a picture
with two-particle excitations out of a new ground state. This was
compared to the entanglement of an effective two-particle state,
obtained by a reduced density matrix approach, applicable for
conductors with arbitrary scattering amplitudes. Moreover, the
qualitative difference between the reduced two-particle density matrix
approach and a projection scheme was discussed.

We showed that the two different approaches, redefinition of the
vacuum state and two-particle state reduction, in general give rise to
different two-particle states and consequently to different
entanglement. This was illustrated by investigating in detail the
orbital entanglement for the simple mesoscopic conductor, focusing on
the spin-polarized regime.  In the tunneling limit, we applied our
approach with the redefinition of the vacuum state to the systems
proposed in Refs. \cite{Sam1}, a normal-superconducting
heterostructure, and in \cite{Sam2}, a conductor in the Quantum Hall
regime. This showed the qualitative as well as quantitative
similarities between the entangled states emitted in the two
systems.

\section*{Acknowledgment}
This work was supported by the Swiss National Science Foundation
and the program for Materials with Novel Electronic Properties.\\


\clearpage

\section*{References}

\begin{thebibliography}{02}
\bibitem{Sam1}  P. Samuelsson, E.V. Sukhorukov, and M. B\"uttiker,
Phys. Rev. Lett. {\bf 91}, 157002 (2003).
\bibitem{Been1} 
C.W.J. Beenakker, C. Emary, M. Kindermann, J.L. van Velsen, Phys. Rev. Lett. {\bf 91}, 147901 (2003).
\bibitem{Sam2} P. Samuelsson, E.V. Sukhorukov, and M. B\"uttiker,
Phys. Rev. Lett. {\bf 92}, 026805 (2004).
\bibitem{Les}
G.B. Lesovik, T. Martin and
G. Blatter, Eur. Phys. J. B {\bf 24}, 287 (2001).
\bibitem{Cht}
N. M. Chtchelkatchev, G. Blatter, G. B. Lesovik, Th. Martin,
Phys. Rev. B {\bf 66}, 161320 (2002).
\bibitem{Faoro} L. Faoro,
F. Taddei, and R. Fazio, Phys. Rev. B {\bf 69}, 125326 (2004).
\bibitem{Been2} 
C.W.J. Beenakker, M. Kindermann, Phys. Rev. Lett. {\bf 92}, 056801
(2004).
\bibitem{Sam3}
M. B\"uttiker, P. Samuelsson and E.V. Sukhorukov, Physica E {\bf 20}, 33 (2003).
\bibitem{Been3}
C.W.J. Beenakker, M. Kindermann, C.M. Marcus, A. Yacoby, {\it Fundamental Problems of Mesoscopic Physics}, eds. I.V. Lerner, B.L. Altshuler, and Y. Gefen, NATO Science Series II. Vol. 154 (Kluwer, Dordrecht, 2004).
\bibitem{Leb1}
 A. V. Lebedev, G. B. Lesovik, G. Blatter, Phys. Rev. B {\bf 71}, 045306 (2005).
\bibitem{Leb2}
 A. V. Lebedev, G. Blatter, C. W. J. Beenakker, G. B. Lesovik,
 Phys. Rev. B {\bf 69}, 235312 (2004).
\bibitem{Naz}
A. Di Lorenzo, Yu. V. Nazarov, cond-mat/0408377.
\bibitem{Schrod}
E. Schr\"odinger, Naturwissenschaften, {\bf 23}, 807 (1935); {\bf 23}, 844 (1935).
\bibitem{EPR}
A. Einstein, B. Podolsky and N. Rosen, Phys. Rev. {\bf 47}, 777
(1935).
\bibitem{Bell}
J.S. Bell, Physics {\bf 1}, 195 (1965); Rev. Mod. Phys. {\bf 38}, 447 (1966).
\bibitem{Gisin} W. Tittel, J. Brendel, H. Zbinden, and N. Gisin, Phys. Rev. Lett {\bf 81}, 3563 (1998).\bibitem{Zeilinger}
G. Weihs, T. Jennewein, C. Simon, H. Weinfurter and A. Zeilinger, Phys. Rev. Lett {\bf 81}, 5039 (1998).
\bibitem{Aspectnat}
A. Aspect, Nature {\bf 398}, 189 (1999).
\bibitem{qinfo}
M. Nielsen and I. Chuang, {\it Quantum Computation and Quantum
Information} (Cambridge Univ. Press, Cambridge, 2000).
\bibitem{quantcryp}
A.K. Ekert, Phys. Rev. Lett. {\bf 67}, 661 (1991).
\bibitem{quanttelep}
C.H. Bennett, G. Brassard, C. Cr\'epeau, R. Josza, A. Peres, and W.K. Wooters, Phys. Rev. Lett. {\bf 70}, 1895 (1993).  
\bibitem{quantcode}
C.H. Bennett and S.J. Wiesner, Phys. Rev. Lett. {\bf 69}, 2881 (1992).
\bibitem{entrevs1}
M. Keyl, Phys. Rep. {\bf 369}, 431 (2002).
\bibitem{entrevs2}
D. Bruss, J. Math. Phys, {\bf 43}, 4237 (2002).
\bibitem{Vedral}
For some recent illuminating examples, see e.g. V. Vedral, quant-ph/0302040.
\bibitem{Peres}
A. Peres, {\it Quantum theory: concepts and methods}, Doordrecht, Kluwer, 1993. \bibitem{Schliemann1}
J. Schliemann, D. Loss, and A. H. MacDonald,  Phys. Rev. B {\bf 63}, 085311
(2002).
\bibitem{Schliemann2}
K. Eckert, J. Schliemann, D. Bruss, and M. Lewenstein, Ann. Phys. {\bf 299}, 88
(2002).
\bibitem{Zanardi} 
P. Zanardi, Phys. Rev. Lett. {\bf 87}, 077901 (2001).
\bibitem{Zanardi2} 
P. Zanardi, D. Lidar, and S. Lloyd, Phys. Rev. Lett. {\bf 92}, 060402 (2004).
\bibitem{Butt2}
M. B\"uttiker, Phys. Rev. B {\bf 46}, 12485 (1992).
\bibitem{Buttrev}
Ya. M. Blanter and M. B\"uttiker, Phys. Rep. {\bf 336}, 1 (2000).
\bibitem{Sukh}
G. Burkhard, D. Loss, and E.V. Sukhorukov, Phys. Rev. B {\bf 89}, 176401
(2000).
\bibitem{Oliver}
X. Maitre, W.D. Oliver, and Y. Yamamoto, Physica E {\bf 6}, 301 (2000).
\bibitem{Carlocomm} 
For works on multiparticle entanglement in mesoscopic systems and detection schemes based on higher order current correlators, see C.W.J. Beenakker, M. Kindermann, Phys. Rev. Lett. {\bf 92}, 056801 (2004) and  C.W.J. Beenakker, C. Emary, M. Kindermann, Phys. Rev. B {\bf 69}, 115320 (2004).
\bibitem{Wiseman1}
S. D. Bartlett, H. M. Wiseman, Phys. Rev. Lett. {\bf 91}, 097903 (2003).
\bibitem{Wiseman2}
H. M. Wiseman, S. D. Bartlett, J. A. Vaccaro, quant-ph/0309046.
\bibitem{Sam6}
P. Samuelsson and M. B\"uttiker, cond-mat/0410581.
\bibitem{Been4}
C.W.J. Beenakker, M. Titov, B. Trauzettel, cond-mat/0502055.
\bibitem{Wooters}
W. Wooters,  Phys. Rev. Lett {\bf 80}, 2245 (1998).
\bibitem{Glauber}
R. Glauber,  Phys. Rev. {\bf 130}, 2529 (1963).
\bibitem{Werner}
R.F. Werner, Phys. Rev. A {\bf 40}, 4277 (1989).
\bibitem{Entcrit1}
A. Peres, Phys. Rev.  Lett {\bf 77}, 1413 (1996).
\bibitem{Entcrit2}
M. Horodecki, P. Horodecki, and  R. Horodecki,  Phys. Lett. A {\bf 223}, 1 (1996).
\bibitem{ShiAlley}
Y.H. Shih and C.O. Alley,  Phys. Rev. Lett. {\bf 61}, 2921 (1988).
\bibitem{Bose}
S. Bose and D. Home, Phys. Rev. Lett. {\bf 88}, 050401 (2002).
\bibitem{Sam4}
 P. Samuelsson, E.V. Sukhorukov, and M. B\"uttiker,
Phys. Rev. B {\bf 70}, 115330 (2004). 
\bibitem{Ou}
Z. Y. Ou, L. J. Wang, X. Y. Zou, and L. Mandel, 
Phys. Rev. A {\bf 41}, 566 (1990). 
\bibitem{Recher}
P. Recher, E. V. Sukhorukov, and D. Loss,
Phys. Rev. B {\bf 63}, 165314 (2001).
\bibitem{Prada}
E. Prada and F. Sols, Eur. Phys. J. B {\bf 40}, 379 (2004).
\bibitem{Recher2}
P. Recher and D. Loss,
Phys. Rev. B {\bf 65}, 165327 (2002).
\bibitem{Bena}
C. Bena, S. Vishveshwara,
L. Balents and M. P. A. Fisher, Phys. Rev. Lett {\bf 89}, 037901
(2002).
\bibitem{Recher3}
P. Recher and D. Loss, Phys. Rev. Lett {\bf 91}, 267003
(2003).
\bibitem{Bouchiat} 
V. Bouchiat, N. Chtchelkatchev, D.Feinberg,
G.B. Lesovik, T. Martin, J. Torres, Nanotechnology {\bf 14}, 77
(2003).
\bibitem{Fein}
O. Sauret, T. Martin, D. Feinberg, cond-mat/0410325.
\bibitem{Datta}
S. Datta, P.F. Bagwell and M.P. Anantram, Phys. Low-Dim. Struct. {\bf
  3}, 1 (1996).
\bibitem{Sam5}
P. Samuelsson, E. V. Sukhorukov and M. B\"uttiker, Turk. J. Phys. {\bf 27}, 481 (2003). 

\end{thebibliography}
\end{document}